\documentclass{an}
\usepackage{times}
\usepackage{graphicx}
\usepackage{verbatim}

\begin{document}


\title{Towards ensemble asteroseismology of the young open clusters $\chi$~Persei and NGC\,6910}

\author{%
S.~Saesen\inst{1} \and
A.~Pigulski\inst{2} \and
F.~Carrier\inst{1} \and
G.~Michalska\inst{2} \and
C.~Aerts\inst{1} \and
J.~De Ridder\inst{1} \and
M.~Bri\-quet\inst{1} \and
G.~Handler\inst{3} \and
Z.~Ko{\l}aczkowski\inst{2} \and
B.~Acke\inst{1} \and
E.~Bauwens\inst{1} \and
P.G.~Beck\inst{1,3} \and
Y.~Blom\inst{1} \and
J.~Blommaert\inst{1} \and
E.~Broeders\inst{1} \and
M.~Cherix\inst{4} \and
G.~Davignon\inst{1} \and
J.~Debosscher\inst{1} \and
P.~Degroote\inst{1} \and
L.~Decin\inst{1} \and
S.~De\-haes\inst{1} \and
W.~De Meester\inst{1} \and
P.~Deroo\inst{1} \and
M.~Desmet\inst{1} \and
R.~Drummond\inst{1} \and
J.R.~Eggen\inst{5} \and
J.~Fu\inst{6} \and
K.~Ga\-zeas\inst{7} \and 
G.A.~Gelven\inst{5} \and
C.~Gielen\inst{1} \and
R.~Huygen\inst{1} \and 
X.~Jiang\inst{8} \and
B.~Kalomeni\inst{9} \and 
S.-L.~Kim\inst{10} \and
D.H.~Kim\inst{10} \and
G.~Kopacki\inst{2} \and
J.-H.~Kwon\inst{10} \and
D.~Ladjal\inst{1} \and 
C.-U.~Lee\inst{10} \and
Y.-J.~Lee\inst{10} \and 
K.~Le\-fever\inst{1,11} \and 
A.~Leitner\inst{3} \and 
P.~Lenz\inst{3,12} \and
A.~Liakos\inst{7} \and 
D.~Lorenz\inst{3} \and 
A.~Narwid\inst{2} \and
P.~Niarchos\inst{7} \and
R.~{\O}stensen\inst{1} \and
E.~Poretti\inst{13} \and
S.~Prins\inst{1} \and
J.~Provencal\inst{14} \and
E.~Puga Antol\'{\i}n\inst{1} \and
J.~Puschnig\inst{3} \and
G.~Raskin\inst{1} \and
M.D.~Reed\inst{5} \and 
M.~Reyniers\inst{1} \and
E.~Schmidt\inst{3} \and
L.~Schmitzberger\inst{3} \and
M.~Spano\inst{4} \and
B.~Steininger\inst{3} \and
M.~St\c{e}\'slicki\inst{2} \and
K.~Uytterhoeven\inst{13,15} \and
J.~Vanautgaerden\inst{1} \and
B.~Vandenbussche\inst{1} \and 
V.~Van Helshoecht\inst{1} \and
E.~Vanhollebeke\inst{1} \and
H.~Van Winckel\inst{1} \and 
T.~Verhoelst\inst{1} \and
M.~Vu\v{c}kovi\'c\inst{1} \and
C.~Waelkens\inst{1} \and
G.W.~Wolf\inst{5} \and
K.~Ya\-kut\inst{1,16} \and
C.~Zhang\inst{6} \and
W.~Zima\inst{1}
}

\institute{
    Instituut voor Sterrenkunde, Katholieke Universiteit Leuven, Leuven, Belgium \and
    Instytut Astronomiczny Uniwersytetu Wroc{\l}awskiego, Wroc{\l}aw, Poland \and
    Institut f\"ur Astronomie, Universit\"at Wien, Wien, Austria \and
    Observatoire de Gen\`eve, Universit\'e de Gen\`eve, Sauverny, Switzerland \and
    Department of Physics, Astronomy and Material Science, Missouri State University, USA \and
    Department of Astronomy, Beijing Normal University, Beijing, China \and
    Department of Astrophysics, Astronomy and Mechanics, University of Athens, Athens, Greece \and
    National Astronomical Observatories, Chinese Academy of Sciences, Beijing, China \and
    Izmir Institute of Technology, Department of Physics, Izmir, Turkey \and
    Korea Astronomy and Space Science Institute, Daejeon, South Korea \and
    Belgian Institute for Space Aeronomy, Brussels, Belgium \and
    Copernicus Astronomical Centre, Warsaw, Poland \and
    INAF - Osservatorio Astronomico di Brera, Merate, Italy \and
    Department of Physics and Astronomy, University Delaware, USA \and
    Laboratoire AIM, CEA/DSM-CNRS-Universit\'e Paris Diderot, Gif-sur-Yvette, France \and
    Department of Astronomy \& Space Sciences, Ege University, Izmir, Turkey
}


\abstract{%
As a result of the variability survey in $\chi$~Persei and NGC\,6910, the number of $\beta$~Cep stars 
that are members of these two open clusters is increased to twenty stars, nine in NGC\,6910 and eleven 
in $\chi$~Persei. We compare pulsational properties, in particular the frequency spectra, of $\beta$~Cep 
stars in both clusters and explain the differences in terms of the global parameters of the clusters.  
We also indicate that the more complicated pattern of the variability among B type stars in $\chi$~Persei 
is very likely caused by higher rotational velocities of stars in this cluster. We conclude that the sample 
of pulsating stars in the two open clusters constitutes a very good starting point for the ensemble 
asteroseismology of $\beta$~Cep-type stars and maybe also for other B-type pulsators.}
\maketitle

\section{Introduction}
From the point of view of seismic modelling, the most important stellar parameters are masses, radii, 
luminosities, and effective temperatures. For single stars these parameters are usually poorly 
determined, especially for distant stars. On the other hand, if pulsators are members of an open cluster, 
the individual stellar parameters can be constrained much better. First, it can be safely assumed that 
the members have the same age, distance and chemical composition. In consequence, one can adopt that 
they are located on the same isochrone.  This helps to pinpoint stellar masses, especially if there 
are eclipsing binaries among cluster members that can be used to yield their masses and radii.

These considerations form a background for an {\it ensemble} asteroseismology of stars
in open clusters, where individual pulsators need no longer be modelled independently. Through the 
location on the isochrone, the masses of stars and consequently their frequency spectra are closely 
related and can be matched simultaneously. 
In general, we would like to point out that this kind of modelling has great potential in terms of 
(i) understanding the pulsations of different groups of stars observed in a cluster, (ii) estimating 
the cluster parameters, and (iii) testing the physics, e.g., the opacities. 

The possibility of ensemble asteroseismology motivated us to carry out multisite campaigns on three 
open clusters that were known to contain $\beta$~Cep-type pulsators, and for which we expected to 
discover several more. The first cluster, NGC\,3293, 
will not be discussed here. Preliminary results of the campaign on this cluster were presented 
by Handler et al.~(2008). The other two clusters we selected are NGC\,884 ($\chi$~Persei) and NGC\,6910.
Before the campaign, two $\beta$~Cep stars and some candidates were known in $\chi$~Persei 
(Krzesi\'nski \& Pigulski 1997, 2000), four in NGC\,6910 (Ko{\l}aczkowski et al. 2004). 

\begin{figure}[!t]
\includegraphics[width=3.2in]{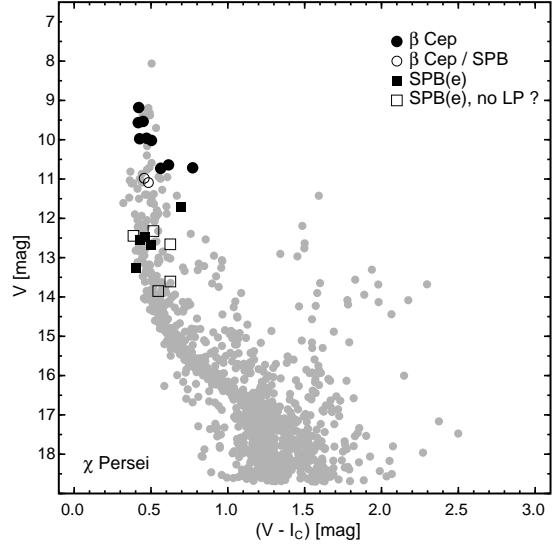}
\caption{Colour-magnitude diagram for $\chi$~Persei. The magnitudes and colours were taken from Keller 
et al.~(2001) and Currie et al.~(2010). The data were corrected for the systematic differences between 
these two sources of photometry. The diagram shows only those stars that were observed during 
the campaign. See text for the explanation of different symbols used to show variable stars.}
\label{chiper-cmd}
\end{figure}

\begin{figure}[!t]
\includegraphics[width=3.2in]{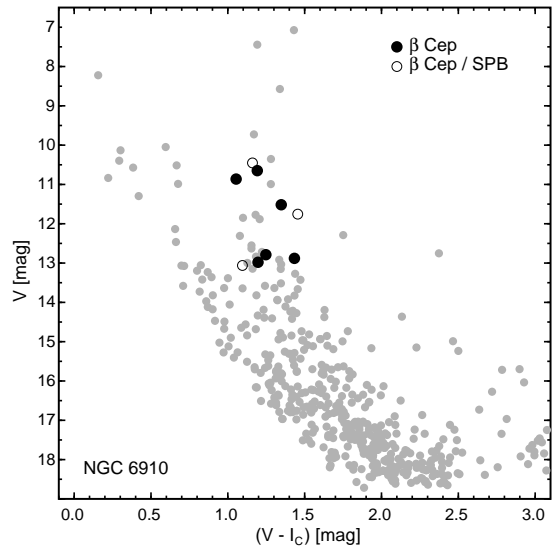}
\caption{Colour-magnitude diagram for NGC\,6910. The photometry was obtained from the campaign data
transformed to the standard system.}
\label{6910-cmd}
\end{figure}

The two open clusters, $\chi$~Persei and NGC\,6910, are different in many ways. The mass of $\chi$~Persei 
is estimated to be 3700~M$_\odot$, the age is about 14~Myr (Currie et al.~2010). NGC\,6910 is about an 
order of magnitude less massive and much youn\-ger; the age of this cluster was estimated to be 
6\,$\pm$\,2~Myr by Ko\-{\l}acz\-kow\-ski et al.~(2004). NGC\,6910 is also more reddened than $\chi$~Persei 
with considerable differential reddening: the range of $E(B-V)$ colour excess amounts to 0.52--0.56~mag 
for $\chi$~Persei and 1.0--1.4~mag for NGC\,6910. The colour-mag\-ni\-tu\-de diag\-rams (CMDs) for 
$\chi$~Persei and NGC\,6910 are shown in Figs.~\ref{chiper-cmd} and \ref{6910-cmd}, respectively. While 
the main sequence of member stars can be clearly seen in the CMD of $\chi$~Persei with only a small 
contamination of field stars, the CMD for NGC\,6910 is severely contaminated. Due to 
the large (and differential) reddening, the cluster main sequence in Fig.~\ref{6910-cmd} is 
smeared and located right of the bluest field stars.

\section{Observations, reductions and analysis}
The campaign was carried out in three seasons, 2005--2007, involving about 70 observers who used 
15 telescopes. For each cluster, over a thousand observing hours were obtained resulting in a 
detection threshold of about 0.2\,--\,0.3~mmag for periodic signals. A detailed description of 
the data obtained for $\chi$~Persei was given by Saesen et al.~(2010). The amount and quality of 
the data for NGC\,6910 is very similar. For a detailed description of the calibration and reduction 
procedures, as well as the results of the variability survey in $\chi$~Persei, we also refer the 
reader to Saesen et al.~(2010). 

The results for NGC\,6910 we are presenting here are based on the analysis of the data from only 
two sites, Bia{\l}\-k\'ow and Xinglong. In terms of the amount and quality of the data, these two 
sites contributed mostly to the final result. The full variability survey for NGC\,6910 will be 
published once all data are reduced and analysed.

\section{Some results and discussion}
Since we discovered many $\beta$~Cep stars, we will focus on the differences in the frequency 
spectra of these stars in both clusters in our short discussion. However, there are other main-sequence 
pulsators in both clusters, in particular slowly pulsating B (SPB) and $\delta$~Sct stars, that 
can also be used in seismic modelling. 

In general, the separation in frequency between p and g modes in massive stars is relatively good 
for non-evolved stars. In the course of main sequence evolution, however, the frequencies of g modes increase 
eventually replacing the frequencies of p modes through avoided crossings. These modes are 
called `mixed' due to their mixed character, g in the interior and p in the envelope. Although 
the occurrence of a p mode (or at least a mixed mode) is a prerequisite for the classification 
of a star as a $\beta$~Cep-type variable, for lack of mode identification we usually use a more 
practical definition of a $\beta$~Cep-type star. For example, early B-type stars showing periodic variations 
with periods shorter than 0.3~d can be termed $\beta$~Cep stars (see, e.g., Sterken \& Jerzykiewicz 1993). 
Following this definition, we classify eleven stars in $\chi$~Persei  and nine in NGC\,6910 as 
$\beta$~Cep stars (see Figs.~\ref{chiper-cmd} and \ref{6910-cmd}). This makes the two clusters 
comparable to four other young open clusters known to be rich in $\beta$~Cep stars: NGC\,3293, 
NGC\,4755, NGC\,6231 and h~Persei (NGC\,869).

\begin{figure}[!t]
\includegraphics[width=3.2in]{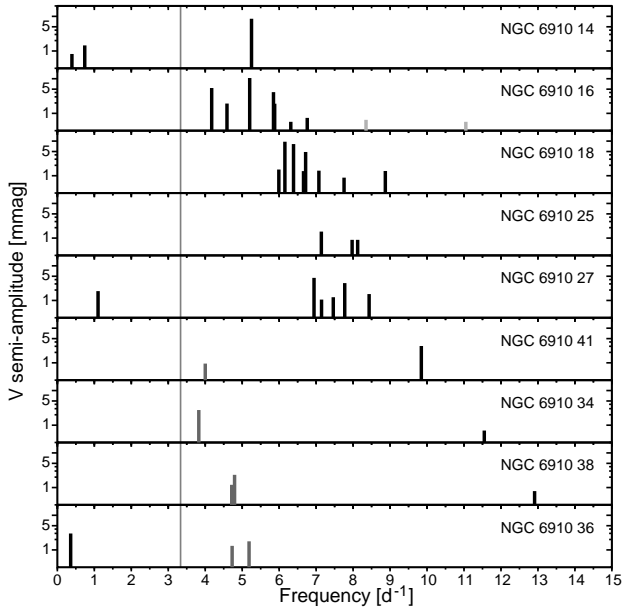}
\caption{Schematic frequency spectra for nine $\beta$~Cep stars in NGC\,6910. The vertical line 
denotes a period equal to 0.3~d. The two light gray bars denote a combination mode and an harmonic. 
See text for the explanation of the dark gray bars.}
\label{6910-frsp}
\end{figure}

\begin{figure}[!t]
\includegraphics[width=3.2in]{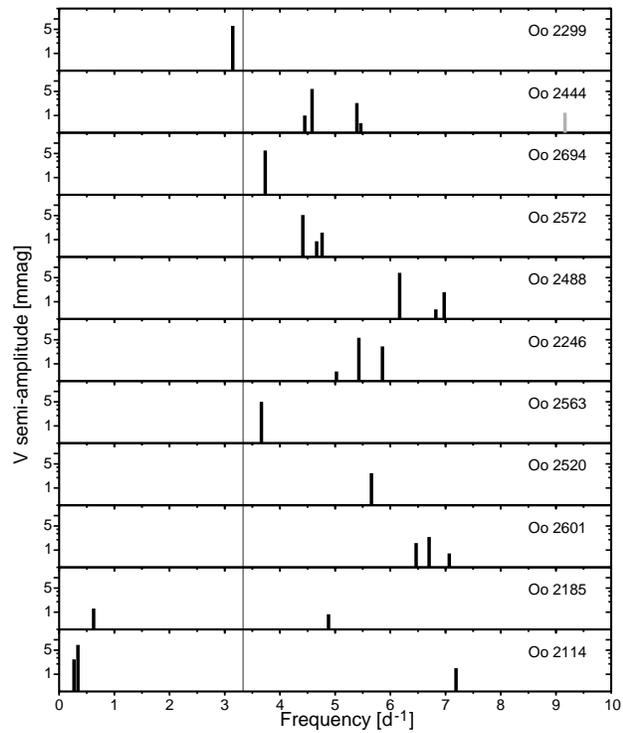}
\caption{The same as in Fig.~\ref{6910-frsp} but for eleven $\beta$~Cep stars in $\chi$~Persei.}
\label{chi-frsp}
\end{figure}

Accounting for differential extinction, we have plotted in Fig.~\ref{6910-frsp} the frequency 
spectra of the nine $\beta$~Cep stars found in NGC\,6910 going from the brightest (top) to the 
faintest (bottom). This is also a sequence of decreasing mass. In addition, because of the same 
age and different masses, the sequence goes from more evolved to less evolved stars. A striking 
feature that can be seen in Fig.~\ref{6910-frsp} is a strip of modes with increasing frequency 
when we go from more massive (bigger) to less massive (smaller) stars. This is exactly what is 
expected for p modes. Additionally, three $\beta$~Cep stars in this cluster, NGC\,6910-14, 27, 
and 36\footnote{The numbers we use follow the Oosterhoff (1937) for $\chi$~Persei and the WEBDA 
(http://www.univie.ac.at/webda) designation for NGC\,6910.}, show the presence of low-frequency 
modes, which can be interpreted as g mo\-des. These stars are therefore good candidate hybrid 
$\beta$~Cep/SPB stars. Moreover, we see that, in the four faintest stars, another group of modes 
occurs (shown with dark gray bars). All they have frequencies in the range between 4 and 5.5~d$^{-1}$. 
Without detailed modelling it is difficult to say if they are p or g modes; f mode(s) are also 
a possibility. 

\begin{figure}[!t]
\includegraphics[width=3.2in]{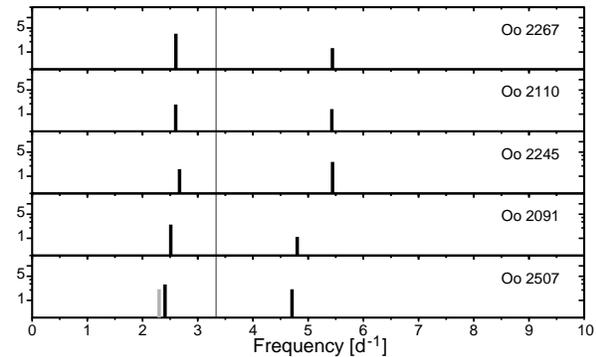}
\caption{The same as in Fig.~\ref{6910-frsp} but for five other stars from $\chi$~Persei showing 
variability with periods shorter than 0.3~d.}
\label{chi-midb}
\end{figure}

The frequency spectra for $\beta$~Cep stars in $\chi$~Persei are presented in Fig.~\ref{chi-frsp}. 
When we go from the most to the least massive $\beta$~Cep stars in this cluster (from top to 
bottom in Fig.~\ref{chi-frsp}), the frequencies of the detected modes do not change as monotonically 
as in NGC\,6910. The most likely explanation of this behaviour is the fast rotation of stars in 
$\chi$~Persei. The projected rotational velocities, $V\sin i$, were measured for many stars in 
$\chi$~Persei and for at least seven $\beta$~Cep stars from our sample the values of $V\sin i$ 
exceed 100~km\,s$^{-1}$ (Strom et al.~2005, Huang \& Gies 2006). Fast rotation results in 
considerable shifts in frequency for non-axi\-sym\-met\-ric modes. Stars in NGC\,6910 have no 
measurements of their rotational velocities, but there is only one Be star known in this cluster, 
whereas in $\chi$~Persei 20 Be stars are known (Keller et al.~2001). This might mean that, on 
average, $\beta$~Cep-type stars in NGC\,6910 rotate much slower than stars of this type in 
$\chi$~Persei. We also note two hybrid stars, Oo\,2114 and 2185, with modes that have frequencies 
below 1~d$^{-1}$, presumably g modes.

\begin{figure}[!t]
\includegraphics[width=3.2in]{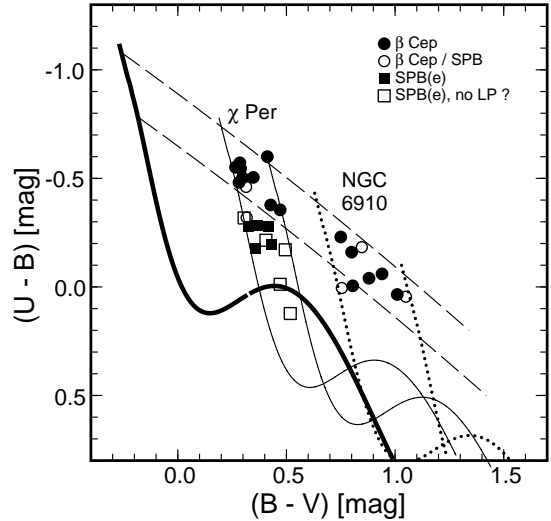}
\caption{The $(U-B)$ vs.~$(B-V)$ diagram for the discussed stars in $\chi$~Persei and NGC\,6910. 
The thick line stands for the relation for unreddened luminosity class V stars (Caldwell et al.~1993).  
The two thin continuous lines denote the same relation shifted by the reddening vectors corresponding 
to $E(B-V)$ = 0.46 and 0.68~mag, a minimum and maximum reddening in $\chi$~Persei. Similar relations 
for NGC\,6910 are shown with dotted lines for $E(B-V)$ = 0.9 and 1.3~mag. Reddening lines, delimiting 
the range of spectral types for $\beta$~Cep stars, are shown with dashed lines.}
\label{ubbv}
\end{figure}

Fast rotation in $\chi$~Persei might also be responsible for the occurrence of several stars with 
periods shorter than 0.3~d which we do not classify as $\beta$~Cep stars because, as can be judged 
from Fig.~\ref{ubbv}, they are mid- to late-B type stars. We divided them into two groups which 
are shown with different symbols in Figs.~\ref{chiper-cmd} and \ref{ubbv}. The first group (shown 
with filled squares) consists of five stars. Their frequency spectra are shown in Fig.~\ref{chi-midb}. 
These stars show bi-periodic variability; the period of one term is shorter than 0.3~d, the period 
of the other, longer. Four of the five stars have measured projected rotational velocities. They 
range from 150 to 250~km\,s$^{-1}$. In addition, Oo\,2091 is known as Be star.  According to many 
different observations, including ground-based data (Uytterhoeven et al.~2007), satellite data 
(Walker et al.~2005, Huat et al.~2009,
Guti\'errez-Soto et al.~2010)
and the surveys in the Magellanic Clouds (Ko\-{\l}acz\-kow\-ski et al.~2006), 
this type of behaviour seems to be very common in fast-rotating stars. Their frequency spectra 
contain usually a large number of terms with frequencies clustering around two values; the ratio 
of these values is usually close to 2.  Although not all five stars in $\chi$~Persei we discussed 
above are known as Be stars, following Walker et al.~(2005), we tentatively classify them as SPB(e) 
stars.

The second group consists of five other variable mid-to late-B type stars, Oo 2752, 2611, 2019, 2406, 
and 2228, shown with open squares in Figs.~\ref{chiper-cmd} and \ref{ubbv}. However, compared to the 
first group, they show only terms with a period shorter than 0.3~d. If due to pulsations, the short periods can 
be explained by fast rotation: rotation can easily shift the frequencies of non-axisymmetric prograde 
g modes to the observed values. It is therefore quite possible that they are also SPB(e) stars in 
which long-period (LP), i.e., low-frequency term(s) are not detected due to low amplitudes.

The results of the variability surveys in $\chi$~Persei and NGC\,6910 which we partially presented 
here form a very good basis for the ensemble asteroseismology in both clusters.  The fast rotation 
for $\chi$~Persei stars is a complication, but, on the other hand, this cluster may help us to 
understand better the interaction between pulsations
and rotation. There are two other open clusters, h Persei and NGC\,3293, with a similar amount of 
data obtained during other campaigns. They are of similar age as $\chi$~Persei and also contain many 
fast-rotating stars.

\acknowledgements
The research leading to these results has received funding from the European
Research Council under the European Community's Seventh Framework Programme
(FP7/2007--2013)/ERC grant agreement n$^\circ$227224 (PROSPERITY), from the
Research Council of K.U.Leuven (GOA/2008/04), from the Fund for Scientific
Research of Flanders (G.0332.06). AP acknowledges the support of the N\,N203 302635 grant from the MNiSzW. 
BA, MB, FC, CG and TV are Postdoctoral Fellows of the Fund for Scientific Research, Flanders (FWO). 
EP and AP acknowledge the support from the European Helio and Asteroseismology Network (HELAS) for the 
participation to the conference.

\end{document}